# Stability assessment of a tailings storage facility using a non-local constitutive model accounting for anisotropic strain-softening

**Mauro G. Sottile**[1,2,*], **Nicolás A. Labanda**[1,2], **Alejandro Kerguelén**[1], **Ignacio A. Cueto**[1] **and Alejo O. Sfriso**[1,2]

[1] SRK Consulting
Chile 300, 1098, Buenos Aires (Argentina)
msottile@srk.com.ar

[2] Engineering Faculty, University of Buenos Aires
Av. Paseo Colón 850, 1063, Buenos Aires (Argentina)

**Abstract** Recent failures of upstream-raised tailings storage facilities (TSF) raised concerns on the future use of these dams. While being cost-effective, they entail higher risks than conventional dams, as stability largely relies on the strength of tailings, which are loose and normally-consolidated materials that may exhibit strain-softening during undrained loading. Current design practice involves limit equilibrium analyses adopting a fully-softened shear strength; while being conservative, this practice neglects the work input required to start the softening process that leads to progressive failure. This paper describes the calibration and application of the NGI-ADPSoft constitutive model to evaluate the potential of static liquefaction of an upstream-raised TSF and provides an indirect measure of resilience. The constitutive model incorporates undrained shear strength anisotropy and a mesh-independent anisotropic post-peak strain softening. The calibration is performed using laboratory testing, including anisotropically-consolidated triaxial compression tests and direct simple shear tests. The peak and residual undrained shear strengths are validated by statistical interpretation of the available CPTu data. It is shown that this numerical exercise is useful to verify the robustness of the TSF design.

## 1 Introduction

Tailings are man-made materials created from mine-rock flour, generally deposited as a viscous mixture into storage facilities (TSFs). The lack of post-deposition compaction and the electrical interaction among finer particles produce loose states, which can be locked by post-deposition early diagenesis (Santamarina, 2019). Upstream raised TSF are highly attractive from the costs point of view, as it minimizes the use of materials other than tailings; however, they are the most vulnerable, as stability largely relies on the strength of the tailings themselves.

Recent upstream-raised TSFs massive failures -such as Merriespruit, Mount Polley, Samarco and Brumadinho (Santamarina, 2019)- have depicted their vulnerability against static liquefaction, which occurs when loose water-saturated tailings undergo a sudden loss of strength due to undrained shearing or by internal fabric collapse. Due to the difficulty of analyzing failure triggering events, international guidelines (e.g. ANCOLD, 2019) recommend to conservatively assume that static liquefaction will occur for brittle/contractive saturated tailings. Thus, current design practice involves limit equilibrium (LE) analyses adopting fully-softened shear strength. While safe, this approach neglects the work required for the material to start a strain-softening process leading to progressive failure, which is acceptable for designing new TSFs but fails short in realistically assess the risk posed by existing TSFs, both operating and closed. In such cases, deformation modelling is called to play.

In this paper, the NGI-ADPSoft constitutive model (Grimstad, 2012) is used to evaluate, by means of a finite element (FE) model, the static liquefaction potential of an upstream-raised TSF after the construction of a reinforcement buttress. The constitutive model incorporates undrained shear strength anisotropy and anisotropic post-peak strain softening; in addition, it uses a regularization technique referred as "over nonlocal strain formulation" to overcome mesh dependency.

## 2 Case study

### 2.1 Geometry

The case study is an upstream-raised TSF with an average external slope of 3.0H:1.0V and a final height of 35 m. The foundation consists of a sandy-clay layer underlain by bedrock. The starter dam and embankment raises are sand-like materials. Tailings are more clay-like and were grouped into three zones based on: i) undrained shear strength; ii) proximity to embankment raises; and iii) age. The TSF is provided with a buttress, built to prevent failure, should static liquefaction occur (Fig 1).

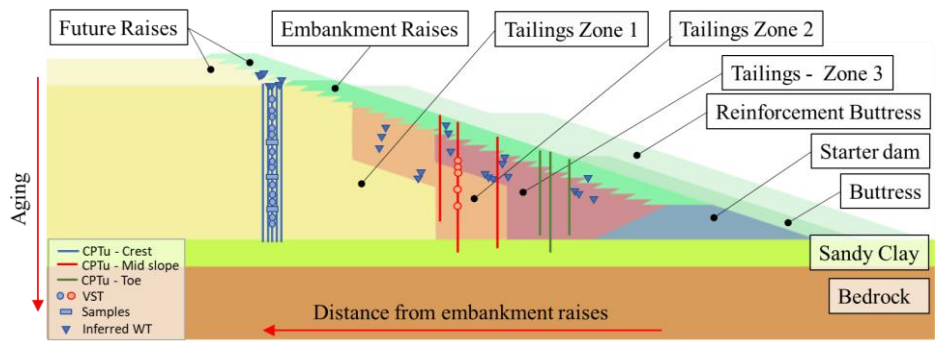

**Fig. 1** Upstream-raised TSF representative cross-section. Material zoning and field-testing location.

# 3 Tailings characterization

## 3.1 General characterization

The behavior of tailings often blurs the common boundaries of 'sand-like' and 'clay-like' materials. As such, inferring material behaviors based on these geological descriptors is not ideal, but usually done to select a framework for stability assessments. In this case, behavior suggests a 'clay-like' material: i) linearity of undrained shear strength with depth from Cone Penetration Tests (CPTu); ii) undrained CPT penetration (i.e. low hydraulic conductivity); iii) undrained shear strength anisotropy, with post-peak softening; iv) no apparent sign of strength recovery after reaching a residual state.

## 3.3 In-situ testing

CPTu soundings including dissipation tests, ball penetrometer tests and Vane Shear Tests (VST) were available. This data was used to estimate the peak and residual undrained shear strength and to identify zones with similar behavioral features.

It was observed that a fully undrained penetration was achieved for all CPTu soundings, as the normalized velocity $V = vD/c_v$ (Finnie & Randolph, 1994) entailed values larger than 30, a limit where partial drainage was experimentally (Randolph & Hope, 2004; Kim et al, 2008) and numerically (Orazalin & Whittle, 2018) proven to have negligible effect on the normalized cone tip resistance. Thus, the peak undrained shear strength can be reliably normalized by the vertical in-situ effective stress, assuming a hydrostatic pore pressure distribution and a submerged tailings unit weight of 12 kN/m$^3$. A log-normal distribution was fitted with the following mean | standard deviation values: i) crest, 0.34 | 0.06; ii) mid-slope, 0.57 | 0.06; ii) toe, 0.57 | 0.20 (Fig 2).

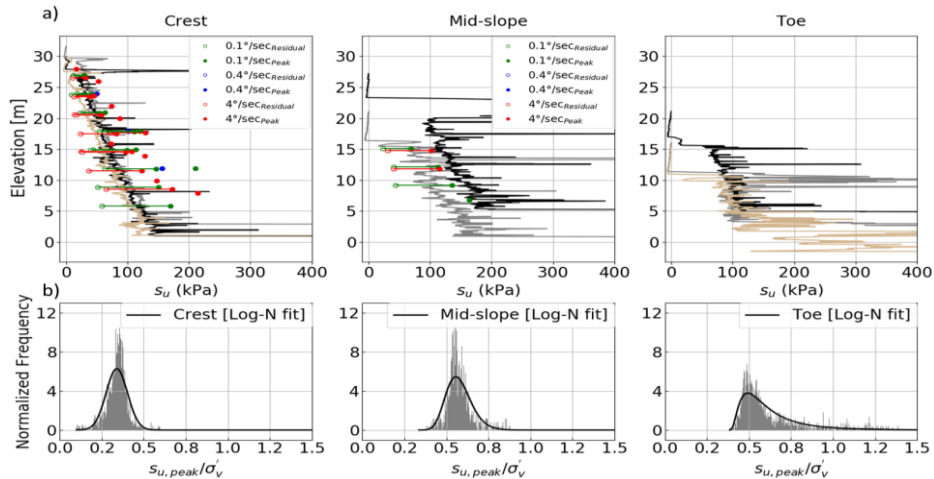

**Fig. 2.** a) CPTu/VST interpreted $s_u$ at crest, mid-slope and toe. b) Log-normal distribution of $s_u/\sigma'_v$.

CPTu data was also used to distinguish tailings behavior based on Robertson (2016): i) updated SBTn, indicating clay-like contractive sensitive (CCS) material at the crest and clay-like/transitional contractive (CC|TC) at the mid-slope and toe; ii) modified normalized small-strain rigidity index ($K_G^*$), that ranges from 200-300 below the crest and 250-330 below the mid-slope, both implying some structuring and showing ageing of the older materials.

Vane Shear Tests were executed at different velocities (0.1°/sec, 0.4°/sec and 4°/sec) with rotations of up to 3600°. Peak and residual undrained shear strengths were defined for rotations of 50° and 3600°, respectively. It was found that: i) peak strength is slightly higher than interpreted from CPTu (Fig. 2a); ii) the ratio between peak and residual strength ranges from 3 to 5; iii) peak strength for 0.1°/sec and 4°/sec are similar, i.e. no partial drainage effect was observed.

Dissipation tests were used to infer the position of the water table. A hydrostatic pore pressure distribution was found and attributed to the low permeability of the foundation.

## 3.2 Laboratory testing

Laboratory tests results are: USCS = CL | CL-ML; Sand | Silt | Clay fraction = 10-35 | 55-70 | 10-20 %; Liquid Limit = 18-24 %; Plasticity Index = 5-10 %; Water Content = 14-26%, Liquidity Index = 0.5-1.4.

Anisotropically consolidated undrained triaxial compression tests (CKUC) and undrained direct simple shear tests (DSS) were performed on undisturbed samples taken from the TSF crest. Samples were consolidated to a vertical effective stress 300 to 600 kPa. Normalized CKUC results are shown in Fig. 3a. Normalized DSS test results are presented in Fig. 3b.

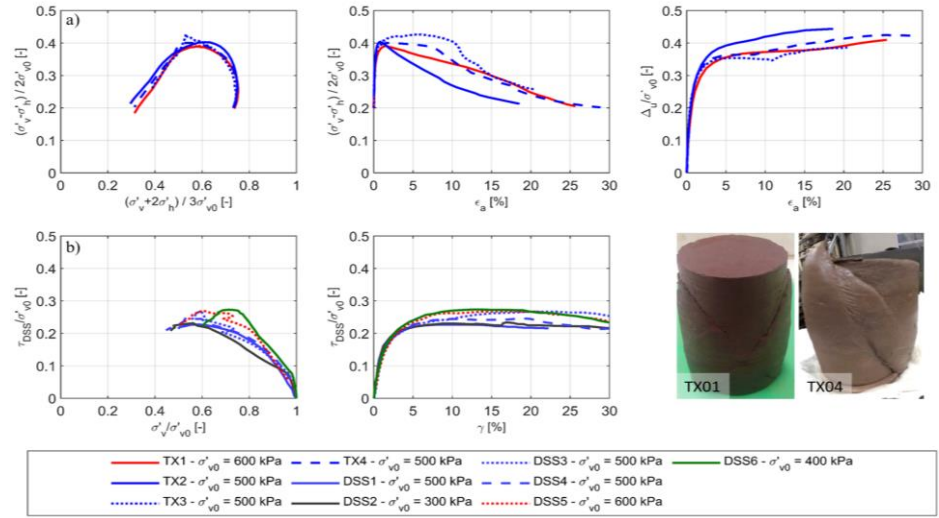

**Fig. 3.** Normalized laboratory test results on tailings undisturbed samples. a) CKUC. b) DSS.

Triaxial tests show: i) $s_{u,TC}/\sigma'_v \sim 0.40$, with strain-at-peak $\varepsilon_{a,peak} \sim 1.0\%$; ii) $s_{ures,TC}/\sigma'_v \sim 0.20$, for strains $\varepsilon_{a,res} \sim 20$-$30\%$; iii) the constant volume friction angle is $\Phi'_{cv} = 37°$; iv) all tests show post-peak softening and shear strain localization in form of shear bands; iv) shear-induced pore pressure is positive, which infer no dominant dilation during shearing -but shear bands were observed. DSS tests show: i) $s_{u,DSS}/\sigma'_v = 0.24$-$0.28$, with strain-at-peak $\gamma_{peak} \sim 5$-$15\%$ ii) moderate post-peak softening occur to $s_{ures,DSS}/\sigma'_v = 0.20$-$0.25$, similar to the TX test results.

## 4 Numerical modelling

### 4.1 *Constitutive model*

The constitutive model NGI-ADPSoft (Grimstad, 2012) is used to evaluate the static-liquefaction performance of the reinforced TSF. The model is formulated on total stresses, using an anisotropic Tresca failure criterion that can capture stress-path-dependent strength along a slip surface and accounts for anisotropic post-peak strain softening. Moreover, it incorporates a regularization technique referred as "over nonlocal strain", which introduces an internal length scale to overcome mesh dependency.

When peak strength is reached at a certain region, local failure and strain localization is initiated and shear bands develop within the initially homogeneous material. After shear banding initiates, further deformation tends to concentrate in these bands. Stress-strain curves are described by: i) peak and residual strengths and their correspondent deviatoric strains for the three modes of shearing ($s_u^A$, $s_u^D$, $s_u^P$, $s_{ur}^A$, $s_{ur}^D$, $s_{ur}^P$, $\gamma^A$, $\gamma^D$, $\gamma^P$, $\gamma_r^A$, $\gamma_r^D$, $\gamma_r^P$); ii) initial shear stress ($\tau_0$); iii) shear modulus at small strain ($G_0$); iv) two shape parameters -$c_1$ and $c_2$- to describe post-peak shape. For more details on the model, the reader is referred to Grimstad (2012) and D'Ignazio (2017).

### 4.2 *Material parameters*

The model, as implemented in Plaxis 2D by the Norwegian Geotechnical Institute (NGI), requires inputs to initialize the parameters and state variables at each of the gauss points within the model geometry. NGI-ADPSoft assumes a linear variation of the undrained shear strength with depth; thus, it requires the input of a reference value ($s_u^A{}_{ref}$) at reference elevation ($y_{ref}$) and its increment with depth ($s_u^A{}_{inc}$); direct simple shear and passive strength are introduced as a ratio to the active strength as $s_u^D/s_u^A$ and $s_u^P/s_u^A$. Residual strength is input in a similar fashion, as a ratio to the peak active strength $s_{u,r}^A/s_u^A$, $s_{u,r}^D/s_u^A$ and $s_{u,r}^P/s_u^A$. For inclined layers, a horizontal coordinate ($x_{ref}$) is defined for the reference elevation, from which an inclination ($\Delta y_{ref}/\Delta x$) is defined.

The initial peak active undrained shear strength ($s_u^A$) contours used is presented in Fig. 4: i) at the crest, $s_u^A{}_{ref} = 1$ kPa at the surface and $s_u^A{}_{inc} = 4.08$ kPa/m, in agreement with the mean $s_u/\sigma'_v$ of 0.34 interpreted from CPTu and assuming $\gamma' = 12$ kN/m$^3$; ii) at the slope below the raises, $s_u^A{}_{ref}$ is redefined considering the raises weight and an inclination for $y_{ref}$, but $s_u^A{}_{inc}$ remains 4.08 kPa/m; ii) for the mid-slope and toe, the $s_u^A{}_{ref}$ is adopted based on an average height of the overlying raises/buttress, and the increase of strength ratio is $s_u^A{}_{inc}$ = 6.84 kPa/m (based on CPTu with $s_u/\sigma'_v$ of 0.57). The ratios $s_u^D/s_u^A$ and $s_u^P/s_u^A$ are adopted 0.75 and 0.50, respectively, based on lab data and Ladd (2003). Residual strength is adopted as $s_{u,r}^A/s_u^A = s_{u,r}^D/s_u^A = s_{u,r}^P/s_u^A$ = 0.30 (i.e. all the same, considering sensitivity ~3.3). Strains at peak are adopted as $\gamma^A|\ \gamma^D|\ \gamma^P$ = 1| 6| 8% and residual strains are adopted as $\gamma_r^A = \gamma_r^D = \gamma_r^P$ = 30 %.

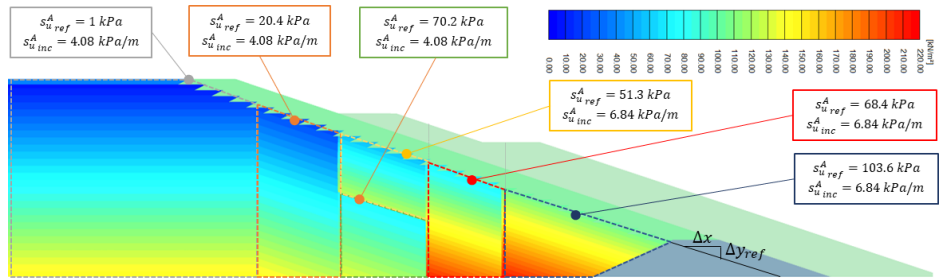

**Fig. 4.** NGI-ADPSoft model. Contours of $s_u^A$ defined for the reinforced TSF configuration.

### 4.3   Geometry and mesh

A 2D numerical model was performed in Plaxis to assess the undrained stability of the raised TSF including the reinforcement buttress. To evaluate mesh-dependency, two meshes were defined: coarse and fine, with 5768 and 11482 15-noded triangular elements, respectively (Fig 5). The model considers: i) water table calibrated from Fig 1; ii) starter dam, raises and buttress are modelled with MC models with Φ'=30 | 32°; iii) clay foundation as HSS undrained material with Φ'=30°; and iv) tailings characterization as described above.

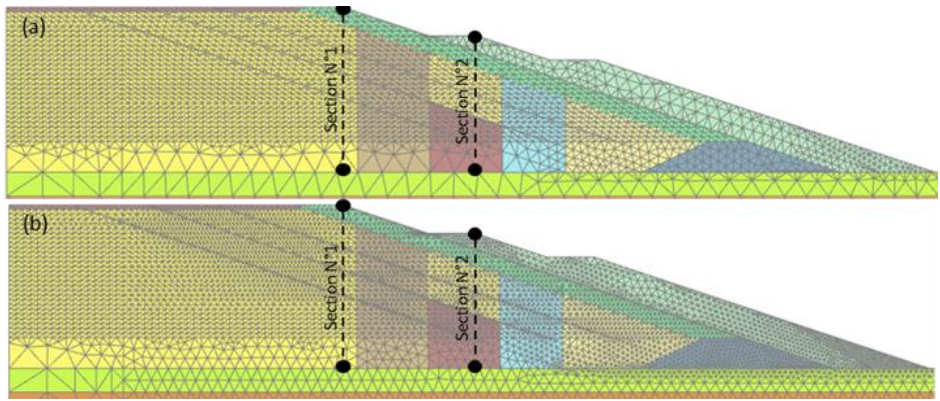

**Fig. 5.** Plaxis 2D model. Mesh and control sections. a) coarse. b) fine.

Material strength parameters were reduced until convergence was no longer achieved. Thus, $s_u^A{}_{ref}$ and $s_u^A{}_{inc}$ were reduced for NGI-ADPSoft, while c' and tan Φ' were reduced for MC and HSS materials manually. Method B shown in Fig. 6 was employed, as it properly accounts for the difference in uncertainty between peak and residual states, implicitly accepted by (ANCOLD, 2019) which requires a Factor of Safety 1.50 for peak strength and a Factor of Safety 1.0 for residual strength.

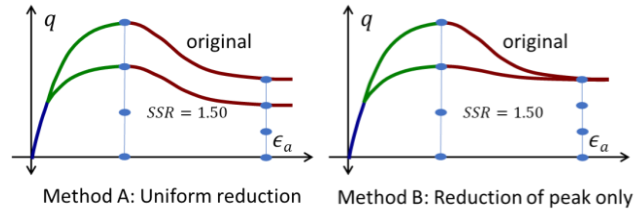

**Fig. 6.** Method A, standard SSR; Method B, recommended for strain-softening materials.

The result of this analysis was SSR = 1.35. For comparison purposes, LE analyses using residual shear strength entailed FoS ~ 1.00. The difference between these two values can be understood as a measure of the work input required to trigger static liquefaction of this particular TSF, expressed in a way that can be readily understood by the industry.

Mesh independency was evaluated by comparing displacements of the two meshes at the two control sections (Fig 7), where red lines represent peak strength parameters and green | blue | black lines represent SSRs 1.20 | 1.30 | 1.35(F) respectively. A good fitting between both meshes is obtained, proving mesh-independent results.

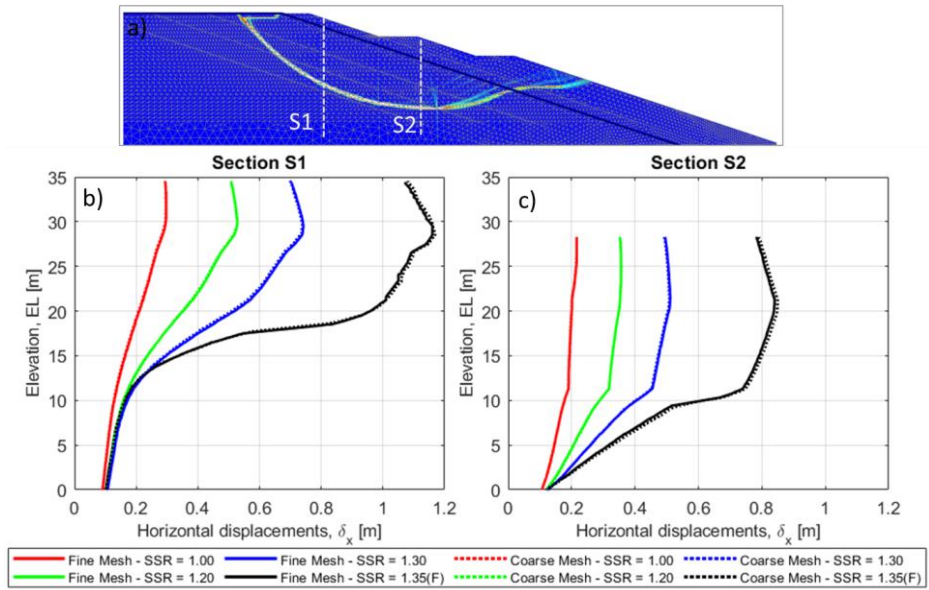

**Fig. 7** a) Failure surface. Horizontal displacements for each SSR at: b) Section N°1, c) Section N°2.

8## 6 Conclusions

The NGI-ADPSoft constitutive model was used to evaluate, by means of a finite element model, the static liquefaction potential of an upstream-raised TSF after the construction of a reinforcement buttress. Calibration was performed using laboratory (CK0UC and DSS); and field test data (CPTu and VST). A procedure where the peak undrained strength is reduced while the residual strength is maintained was employed. A shear strength reduction parameter SSR = 1.35 was obtained, while LE analyses adopting residual shear strength produced FoS ~ 1.00. The difference between these two valued is considered an indirect estimate of the resilience against static liquefaction, an aspect which plays a vital role in evaluating the robustness of existing TSFs. In addition, the mesh independency of the results was demonstrated by using two different meshes and favourably comparing the output of both, in terms of total displacements at two control sections.